\documentstyle[11pt,moriond,epsfig,psfig]{article}

\def\be{\begin{equation}}
\def\ee{\end{equation}}
\def\bea{\begin{eqnarray}}
\def\eea{\end{eqnarray}}

\begin{document}

\title{PION TRANSITION FORM FACTOR: INTERPLAY OF HARD AND SOFT LIMITS.}
\author{A. E. DOROKHOV}

\vspace*{4cm}

\address{ Bogoliubov Laboratory for Theoretical Physics,
Joint Institute for Nuclear Research,\\ 141980, Dubna, Russia}
\maketitle
\abstracts{
The behavior of the pion transition form factor for the processes $\gamma
^{\star}\gamma \rightarrow \pi ^{0}$ and $\gamma ^{\star }\gamma
^{\star}\rightarrow \pi ^{0}$ at space-like values of photon momenta
is estimated within the nonlocal covariant quark-meson model.
The nonlocal contributions are important to install the axial anomaly
at zero virualitites and give the contributions to the twist-4 power 
correction to the form factor at large virtualities. The leading and next-to-leading
order power asymptotics of the form factor and
the relation between the light-cone pion distribution
amplitudes and the dynamically generated quark mass function are found.}

The pion form factor $M_{\pi
^{0}}(q_{1}^{2},q_{2}^{2},p^{2}) $ for the transition processes $\gamma
^{\star }(q_{1})\gamma (q_{2})\rightarrow \pi ^{0}(p)$ and $\gamma ^{\star
}(q_{1})\gamma ^{\star }(q_{2})\rightarrow \pi ^{0}(p)$, where $q_{1}$ and $%
q_{2}$ are photon momenta, related to the fundamental properties of QCD dynamics at
low and high energies. At zero photon virtualities, the observed two-photon $%
\pi _{0}-$decay width is given by 
\begin{equation}
\Gamma (\pi ^{0}\rightarrow \gamma \gamma )=\frac{m_{\pi _{0}}^{3}}{64\pi }%
M_{\pi ^{0}}^{2}\left( 0,0,m_{\pi _{0}}^{2}\right) =7.78(56)\quad {\rm eV,} 
\end{equation}
and theoretically is related to the chiral anomaly for $\pi _{0}$
\begin{equation}
M_{\pi ^{0}}\left( 0,0,0\right) = (4\pi^2f_\pi)^{-1}. 
\end{equation}
At asymptotically large photon virtualities, the behavior of the form
factor is predicted by perturbative QCD (pQCD). The leading momentum power
dependence is dictated by the scaling property of the pion wave function.
But the coefficients of the asymptotic expansion depend crucially on the internal pion
dynamics, which is parametrized by the nonperturbative pion distribution
amplitudes (DA), $\varphi _{\pi }(x)$, with $x$ being the fraction of the
pion momentum, $p$, carried by a quark.

The existing experimental data from CELLO \cite{CELLO} and CLEO \cite{CLEO}
Collaborations on the form factor $T_{\pi ^{0}}$ for one photon being almost
real, $q_{2}^{2}\approx 0$, with the virtuality of the other photon scanned
up to $8$ GeV$^{2}$, can be fitted by a monopole form factor: 
\begin{equation}
\left. M_{\pi ^{0}}(q_{1}^{2}=-Q^{2},q_{2}^{2}=0)\right| _{fit}=\frac{g_{\pi
\gamma \gamma }}{1+Q^{2}/\Lambda _{\pi }^{2}},\ \ \ \ \ \Lambda _{\pi
}\simeq 0.77\ GeV,  \label{Fpiggfit}
\end{equation}
where $g_{\pi \gamma \gamma }=0.275$ {\rm GeV}$^{-1}$ \ is the two-photon
pion decay constant. The large $Q^{2}$ behavior of the form factor is in
agreement with the lowest order perturbative QCD~prediction \cite{BrLep79} 
\begin{equation}
\left. M_{\pi ^{0}}(q_{1}^{2},q_{2}^{2})\right| _{Q^{2}\rightarrow \infty}=
J^{(2)}\left( \omega \right) \frac{1}{Q^{2}}+
J^{(4)}\left( \omega \right) \frac{1}{Q^{4}}+
O(\frac{\alpha _{s}}{\pi })+O(\frac{1}{Q^{6}}),  \label{AmplAsympt}
\end{equation}
where the leading and next-to-leading order asymptotic coefficients $J\left( \omega \right) $ 
are expressed in terms of
the light-cone pion DAs, $\varphi_{\pi }(x)$: 
\begin{equation}
J^{(2)}\left( \omega \right) =\frac{4}{3}f_{\pi }\int_{0}^{1}
\frac{dx\varphi^{(2)} _{\pi }(x)}{1-\omega ^{2}(2x-1)^{2}},\ \ \ \
J^{(4)}\left( \omega \right) =\frac{8}{3}f_{\pi }\Delta^2\int_{0}^{1}
\frac{dx[1+\omega ^{2}(2x-1)^{2}]\varphi^{(4)} _{\pi }(x)}{[1-\omega ^{2}(2x-1)^{2}]^2}.  \label{J}
\end{equation}
In above expressions, $Q^{2}=-(q_{1}^{2}+q_{2}^{2})\geq 0$ is the total
virtuality of the photons, $\omega
=(q_{1}^{2}-q_{2}^{2})/(q_{1}^{2}+q_{2}^{2})$ is the asymmetry in their
distribution. The distribution 
amplitudes are normalized by $\int_{0}^{1}dx\varphi _{\pi }(x)=1$, $f_{\pi }=92.4$ {\rm MeV} is 
the weak pion decay constant and the parameter $\Delta^2$ will be specified below. 
The first perturbative correction to the leading term has been found in 
\cite{Braaten} and the next-to-leading power correction has been recently discussed
in \cite{twist4} within the collinear operator product expansion.

In this talk, we shall present a nonperturbative approach to calculation of the pion
transition form factor in the total kinematical region. The calculations are 
consistent with the chiral anomaly and gives the leading order and next-to-leading 
order power behaviour in the asymptotic region.
The covariant nonlocal low-energy models, based on the Schwinger-Dyson (SD)
approach to dynamics of quarks and gluons, have many attractive features, as
the approach preserves the gauge invariance and consistent with the low-energy theorems. Furthermore, the
intrinsic nonlocal structure of the model may be motivated by fundamental
QCD processes like the instanton and gluon exchanges. The effective
quark-pion dynamics in separable approximation may be summarized in terms 
of the dressed quark propagator \cite{ADoLT00}
\[
S^{-1}\left( p\right) =\widehat{p}-M\left( p^2\right) , 
\]
the quark-pion vertex 
\[
\Gamma _{\pi }^{a}\left( k,p,k^{\prime }=k+p\right) =
\frac{i}{f_{\pi }}F(k,k^{\prime })\gamma _{5}\tau^{a}, 
\qquad F\left( k,k^{\prime }\right) = 
\sqrt{M\left( k^2\right) M\left( k^{\prime 2}\right)}
\]
and quark-photon vertex 
\[
\Gamma ^{\mu }\left( k,q,k^{\prime }=k-q\right) =eQ\left[ \gamma _{\mu
}-\left( k+k^{\prime }\right) _{\mu }G\left( k,k^{\prime }\right) \right]
,\qquad G\left( k,k^{\prime }\right) =\frac{M\left( k^{\prime 2}\right)
-M\left( k^2\right) }{k^{\prime 2}-k^{2}}, 
\]
where $M(k^{2})$ is the dynamical quark mass. The dynamical mass characterizes 
the momentum dependence
of an order parameter for spontaneous chiral-symmetry breaking and can be
expressed in terms of the nonlocal quark condensate \cite{ADWB01}. The inverse size of the
nonlocality scale, $\Lambda $, is naturally related to the average
virtuality of quarks that flow through the vacuum, $\lambda _{q}^{2}\approx
\Lambda ^{2}$. The value of $\lambda _{q}^{2}$ is known from the QCD sum
rule analysis $\lambda _{q}^{2}$ $\approx 0.5\pm 0.1{\rm GeV}^{2}$ and,
within the instanton model, may be expressed through the average instanton
size, $\rho _{c}$, as $\lambda _{q}^{2}\approx 2\rho _{c}^{-2}$ \cite{DEM97}. 
The pion weak decay constant is expressed by the Pagels-Stokar formula 
\begin{equation}
f_{\pi }^{2}=\frac{N_{c}}{4\pi ^{2}}\int_{0}^{\infty }du\frac{uM(u)\left[
M(u)-uM^{\prime }(u)/2\right] }{\left( u+M^{2}(u)\right) ^{2}},  \label{f_pi}
\end{equation}
where $M^{\prime }(u)=\frac{d}{du}M(u)$.

The invariant $\gamma ^{\ast }\gamma ^{\ast }\pi ^{0}$ amplitude is given by 
\[
{\cal M}\left( \gamma ^{\ast }\left( q_{1},\epsilon _{1}\right) \gamma
^{\ast }\left( q_{2},\epsilon _{2}\right) \rightarrow \pi ^{0}\left(
p\right) \right) =e^{2}\varepsilon _{\mu \nu \rho \sigma }\epsilon _{1}^{\mu
}\epsilon _{2}^{\nu }q_{1}^{\rho }q_{2}^{\sigma }M_{\pi ^{0}}\left(
q_{1}^{2},q_{2}^{2},p^{2}\right) ,
\]
where $\epsilon _{i}^{\mu }(i=1,2)$ are the photon polarization vectors. In
the nonlocal covariant model one finds the contributions of the triangle
diagrams to the invariant amlitude as
\begin{equation}
{\cal M}\left( \gamma_1 ^{\ast } \gamma_2^{\ast } \rightarrow \pi ^{0}\right) =
-\frac{N_{c}}{3f_{\pi }}\int \frac{d^{4}k}{(2\pi )^{4}}
F(k_{+},k_{-})
\left\{ tr\{i\gamma _{5}S(k_{-}){\hat{e}}_{2}S[k-q/2]{%
\hat{e}}_{1}S(k_{+})\}+\right. 
\end{equation}
\[
+tr\{i\gamma _{5}S(k_{-})S[k-q/2]{\hat{e}}_{1}S(k_{+})\}\left( \epsilon
_{2},2k-q_{1}\right) F\left( k-q/2,k_{-}\right) 
\]
\[
\left. +tr\{i\gamma _{5}S(k_{-}){\hat{e}}_{2}S[k-q/2]S(k_{+})\}\left(
\epsilon _{1},2k+q_{2}\right) F\left( k_{+},k-q/2\right) \right\} +\left(
q_{1}\leftrightarrow q_{2};\epsilon _{1}\leftrightarrow \epsilon _{2}\right)
,
\]
where $p=q_{1}+q_{2},$ $q=q_{1}-q_{2},$ $k_{\pm }=k\pm p/2$. In the chiral
limit $\left( p^{2}=m_{\pi }^{2}=0\right) $ with both photons real $\left(
q_{i}^{2}=0\right) $ one finds the result 
\[
M_{\pi ^{0}}\left( 0,0,0\right) =\frac{N_{c}}{6\pi ^{2}f_{\pi }}\int_{0}^{\infty }du%
\frac{uM(u)\left[M(u)-2uM^{\prime }(u)\right]}{\left( u+M^{2}\left( u\right) \right) ^{3}}
 =\frac{1}{4\pi ^{2}f_{\pi }},
\]
consistent with the chiral anomaly. 

The leading behavior
of the form factor at large photon virtualities is given by the contribution
of the first term in (7) and next-to-leading power correction is generated
by the second and third terms in (7). Thus,
for large $q_{1}^{2}=q_{2}^{2}=-Q^{2}/2$ and $p^{2}=0$ the form factor has
the asymptotics 
\begin{equation}
\left. M_{\pi ^{0}}\left( q_{1}^{2}=q_{2}^{2},0\right) \right|
_{Q^{2}\rightarrow \infty }=
\frac{4f_{\pi }}{3Q^{2}}\left(1+\frac{2\Delta^2}{Q^{2}}\right),\ \ 
\Delta^2=\frac{N_c}{4\pi^2 f_\pi^2}\int_{0}^{\infty }du\frac{u^2M^2(u)}
{\left( u+M^{2}\left( u\right) \right) ^{2}}
\end{equation}
which is in agreement with the expressions (4), (\ref{J})  
for the asymptotic coefficients at $\omega =0$. The parameter
$\Delta^2$ has an extra power of $u$ in the integral with respect to (\ref{f_pi})
and is proportional to the matrix element 
$\left\langle \pi(p)\left| g_s\overline{d} 
\tilde{G}_{\alpha\mu}\gamma_\alpha p_\mu u\right| 0\right\rangle$.

In general case at
large $Q^{2}$ the model calculations reproduce the QCD factorization result 
(\ref{AmplAsympt}),(\ref{J}) with the DAs given by 
\begin{equation}
\varphi _{\pi }^{(2)}(x)=\frac{N_{c}}{4\pi ^{2}f_{\pi }^{2}}
\int_{-\infty }^{\infty }\frac{d\lambda }{2\pi }\int_{0}^{\infty }du 
\frac{\left[xM^{3/2}\left( u+i\lambda \overline{x}\right) M^{1/2}\left( u-i\lambda
x\right) +\left( x\leftrightarrow \overline{x}\right) \right]}
{\left( u-i\lambda x+M^{2}\left( u-i\lambda x\right) \right)
\left( u+i\lambda \overline{x}+M^{2}\left( u+i\lambda \overline{x}\right)\right) } ,
\label{WF_VF2}\end{equation}
\begin{equation}
\varphi _{\pi }^{(4)}(x)=\frac{1}{\Delta^{2}}\frac{N_{c}}{4\pi ^{2}f_{\pi }^{2}}
\int_{-\infty }^{\infty }\frac{d\lambda }{2\pi }\int_{0}^{\infty }du 
\frac{u\left[M^{3/2}\left( u+i\lambda \overline{x}\right) M^{1/2}\left( u-i\lambda
x\right) +\left( x\leftrightarrow \overline{x}\right) \right]}
{\left( u-i\lambda x+M^{2}\left( u-i\lambda x\right) \right)
\left( u+i\lambda \overline{x}+M^{2}\left( u+i\lambda \overline{x}\right)\right) } .
\label{WF_VF4}\end{equation}

In Fig. 1 the normalized by unity leading order and next-to-leading order
pion DA are illustrated in comparison with asymptotic DA. For the numerical analysis, 
the dynamical mass function is chosen in the Gaussian
form $M(k^2)=M_q \exp{(-2 k^2/\Lambda^2)},$ where we take $M_q=350$ MeV and fix 
$\Lambda = 1290$ MeV from Eq. (\ref{f_pi}).
Then the value $\Delta^2=0.29$ GeV$^2$ is obtained.
As it is clear from Fig. 1, the model pion DAs at the
realistic choice of the parameters are close to the
asymptotic DA, that is in agreement with earlier works \cite{ADT99pigg}. 
In Fig. 2, for the process $\gamma \gamma ^{\ast
}\rightarrow \pi ^{0}$ ($\omega =1$), we plot the form factor 
$F_{\pi\gamma}(Q^2)=M_{\pi ^{0}}\left( -Q^{2},0\right)$ multiplyed by square 
momentum $Q^2$. In this figure, we indicate the CLEO data and model
predictions for the full form factor and twist-4 contribution to it. 
In the full model form factor the perturbative $\alpha_s-$ correction to the 
leading twist-2 term is effectively taken into account with the running coupling 
that has zero at zero momentum \cite{VAV02}. With such behaviour in the infrared 
region the
perturbative corrections do not influent the chiral anomaly. At high momentum
squared the leading perturbative correction provides negative contribution to 
the form factor and compensate the $Q^{-4}$ correction in the region $2-10$ GeV$^2$.
The unknown perturbative correction to the twist-4 contribution is considered as
inessential.

In conclusion, within the covariant nonlocal model describing the quark-pion
dynamics, we obtain the $\pi \gamma ^{\ast }\gamma ^{\ast }$ transition form
factor in the region up to moderately high momentum transfer squared, where the perturbative
QCD evolution does not yet reach the asymptotic regime. 
From the comparison of the kinematical dependence
of the asymptotic coefficients of the transition pion form factor, as it is
given by pQCD and the nonperturbative model, the relations 
Eqs. (\ref{WF_VF2}, \ref{WF_VF4}) between the pion DAs and the
dynamical mass function are derived. The other possible sources of contributions
to the form factor are from inclusion into the model of vector and axial-vector
mesons. In general these contributions to the pion transition form factor are small.
They do not contribute to the result given by the chiral anomaly for the two-gamma
pion decay. The contribution of the vector mesons to the leading
order asymptotics of the form factor is expected to be small, 
but it may be more important when one treats the twist-4 power corrections. 

{\it Acknowledgments}

I am grateful to N.I. Kochelev, S.V. Mikhailov, Lauro Tomio, M.K. Volkov and
V.L. Yudichev for many useful discussions on topics related to this talk. 
The work is supported by RFBR Grants 01-02-16231 and 02-02-16194 and by Grant
INTAS-2000-366. 
I wish to cordially thank the organizers, in particular J. Tran Thanh Van, E. Auge and
G. Korchemsky, for their hospitality and for creating the perfect working
atmosphere at the Moriond 2002 Conference.

\begin{figure}[tbp]
\vskip -0.5cm \centering
\begin{minipage}[c]{7cm}
\epsfig{file=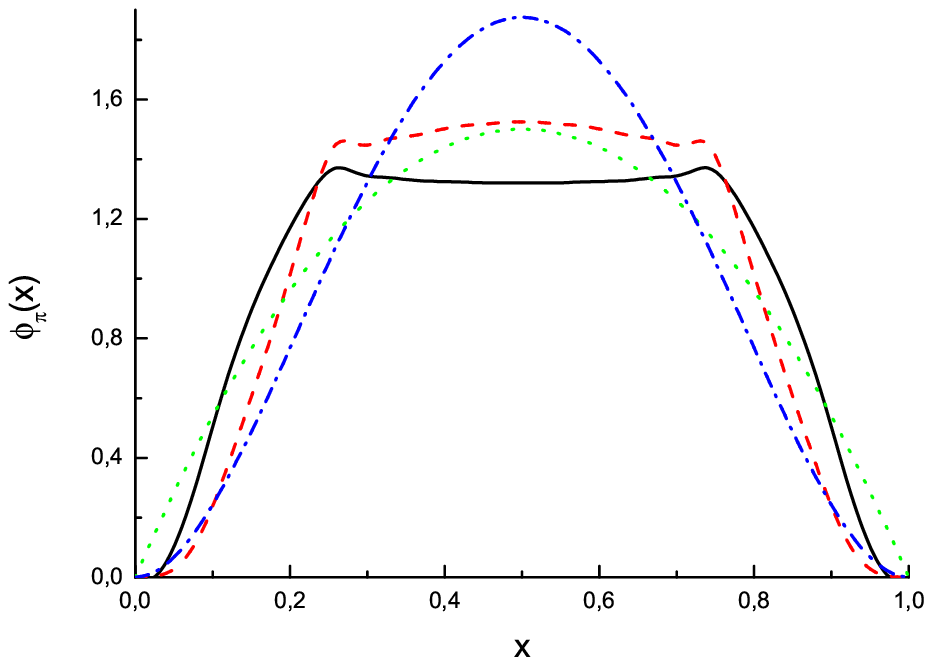,width=6.5cm,height=5.5cm}
\caption{\it  The pion distribution amlitudes (normalized by unity):
the model predictions for twist-2 (solid line) and twist-4 (dashed line) components 
and asymptotic limits of twist-2  (dotted line) and twist-4  (dash-dotted line) amplitudes.}
\label{relx}
\end{minipage}
\hspace*{0.5cm} 
\begin{minipage}[c]{7cm}
\epsfig{file=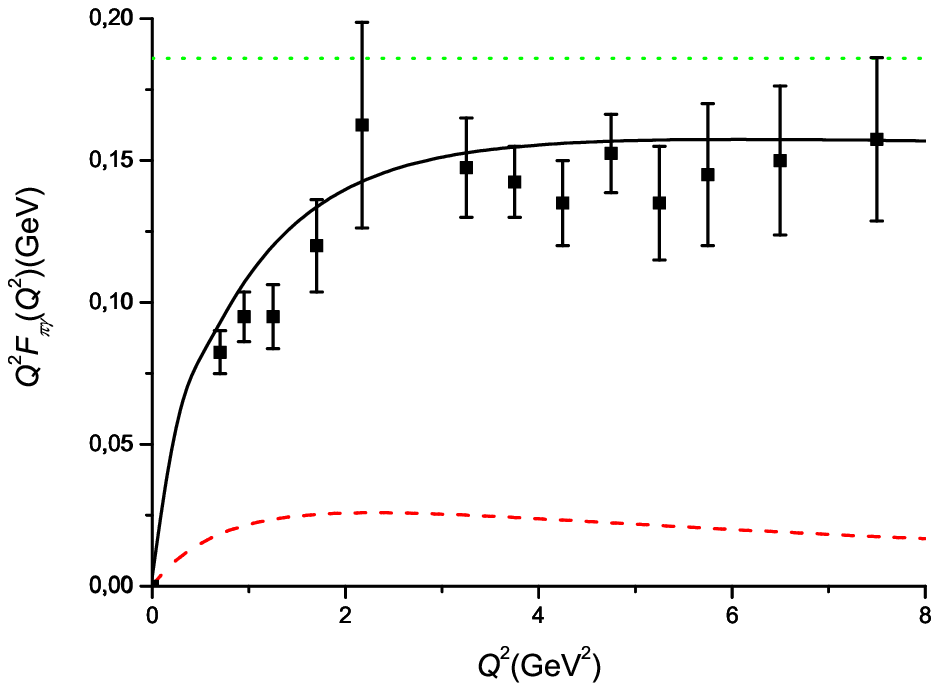,width=6.5cm,height=5.5cm}
\caption{\it The pion-photon transition form factor $Q^2F_{\pi\gamma}(Q^2)$:
the full model prediction (solid line), the twist-4 contribution (dashed line)
and perturbative $2f_\pi$ limit (dotted line). The experimental points are taken
from $^2$.}
\label{rels}
\end{minipage}
\end{figure}


\end{document}